\documentclass[11pt]{article}
\usepackage{amsmath}  
  
\font\titlefont=cmbx10 scaled \magstep3  
\input{tcilatex}  
\input{epsf}  
  
\begin{document}  
  
\begin{center} 
{\titlefont The Energy Density in the Casimir Effect}  
\vskip .3in 
V. Sopova\footnote{email: svasilka@tufts.edu} and  
L.H. Ford\footnote{email: ford@cosmos.phy.tufts.edu} \\ 
\vskip .1in 
Institute of Cosmology, 
Department of Physics and Astronomy\\ 
Tufts University\\ 
Medford, Massachusetts 02155\\ 
\end{center} 
  
\begin{abstract}  
We compute the expectations of the squares of the electric and magnetic  
fields in the vacuum region outside a half-space filled with a uniform  
dispersive dielectric. We find a positive energy density of the  
electromagnetic field which diverges at the interface despite the inclusion  
of dispersion in the calculation. We also investigate the mean squared fields 
and the energy density in the vacuum region between two parallel half-spaces. 
Of particular interest is the sign of the energy density. We find that  
the energy density is described by two terms: a negative position  
independent (Casimir) term, and a positive position dependent term with a  
minimum value at the center of the vacuum region. We argue that in some  
cases, including physically realizable ones, the negative term can dominate in a given  
region between the two half-spaces, so the overall energy density can be  
negative in this region.  
\end{abstract}  
 \vskip .1in 
 PACS categories: 12.20.Ds, 03.70.+k, 77.22.Ch,  04.62.+v. 
 
\section{\protect\bigskip\ \ Introduction\ \ \ \ \ \ \ \ \ \ \ \ \ \ }  
 
In 1948 Casimir made the remarkable prediction that there is an attractive 
force between a pair uncharged parallel plane perfect conductors \cite{Casimir}. 
Furthermore, he argued that this force arises solely from a shift in the  
energy of the vacuum state of the quantized electromagnetic field. An early 
attempt by Sparnaay \cite{Sparnaay} to observe this force was inconclusive, 
but in recent years several new experiments \cite{Lamoreaux,Mohideen,Mohideen2, 
Chan,Bressi} have been performed which seem to give good agreement with  
Casimir's prediction. (To be more precise, most of these experiments actually 
measure the force between a plate and a sphere and incorporate a theoretical 
correction to compare to Casimir's result. Of the recent experiments, only 
that of Bressi {\it et al} \cite{Bressi} uses two parallel plates.) 
 
If the energy of the vacuum state is zero in the limit of infinite plate 
separation, then the attractive force found by Casimir would seem to imply 
a negative vacuum energy at finite separation. In fact,  
Brown and Maclay \cite{Brown} showed that  
for perfectly conducting plates, one has a constant negative vacuum energy  
density. This conclusion is of great theoretical interest, because negative 
energy density has the potential to cause some rather bizarre effects in 
gravity theory. (See, for example, Ref.~\cite{FR01} and references therein.) 
However, questions have been raised as to whether the negative energy density 
will still arise in a more realistic treatment in which the plates are not 
perfect conductors \cite{Lang,lam}. In particular, Helfer and Lang \cite{Lang} 
calculated the energy density outside of a single half-space filled with 
a nondispersive dielectric material and obtained a positive result. They 
interpreted this as a positive self-energy density associated with a single 
plate which would add to the negative interaction energy density between 
a pair of plates. Helfer and Lang conjecture that the net Casimir energy  
density might be positive when the self energy is accounted for. If this 
conjecture is correct, then the situation would be analogous to that of 
the energy density in classical electrostatics. A pair of oppositely charged 
particles have a negative interaction energy, but the net energy density, 
which is proportional to the square of the electric field, is always 
positive.  
 
However, the Helfer and Lang calculation does not include dispersion,  
which is essential in a realistic treatment. Numerous authors, beginning 
with Lifshitz \cite{Lif}, have studied the effects of dispersion upon 
Casimir forces. However, these authors have been concerned with the force 
or the total energy, and not the local energy density. The purpose of this  
paper is to present a calculation of the Casimir energy density in a model 
in which dispersion is included. For this purpose, we will use the methods  
of source theory developed by  Schwinger and coworkers \cite{one,sources}.  
This is a method based upon the calculation of Green's functions which is 
especially well suited to dissipative materials, and was used by  
Schwinger {\it et al} \cite{one} to rederive the results of Lifshitz. 
Milonni and Shih \cite{MS92} have used conventional quantum electrodynamics 
to reproduce some of the results of source theory. There has also been  
considerable interest in recent years in quantization of the electromagnetic 
field inside dissipative materials using operator methods \cite{HB92,GW,Mat}. 
The relation between the results of the latter set of authors and those 
of Schwinger {\it et al} has not yet been clarified. 
 
The outline of this paper is as follows: In Sect.~\ref{sec:GF} we review the 
source theory approach as applied to parallel interfaces of dielectric media. 
In Sect.~\ref{sec:one} we compute the  
expectation values of the squares of the electric and magnetic fields in the  
vacuum region outside a half-space filled with a uniform dispersive dielectric. 
We extend this calculation to the case of two parallel dielectric half-spaces 
and also discuss the energy density in Sect.~\ref{sec:two}. Conclusions are 
given in Sect.~\ref{sec:final}.

\section{Green-Function Approach for Multilayer Dielectrics}  
\label{sec:GF} 
 
This section is a review of the formalism of Schwinger {\it et al} \cite{one}.  
One begins by writing the Maxwell equations for the macroscopic  
electromagnetic fields produced by an external polarization source $\mathbf{P%
}$, which formally describes the zero point fluctuations of the fields%
\footnote{  
Heaviside-Lorentz units with $c=\hbar =1$ will be used in this paper.  
Also, it is assumed that the magnetic permeability is unity.}  
  
\begin{gather}  
\mathbf{B}=\nabla \times \mathbf{A},  \notag \\  
\mathbf{E}=-\dot{\mathbf{A}}-\nabla \phi ,  \notag \\  
\nabla \times \mathbf{B}=\epsilon \dot{\mathbf{E}}+\dot{\mathbf{P}},  
\label{max} \\  
\nabla \cdot (\epsilon \mathbf{E}+\mathbf{P})=0,  \notag  
\end{gather}  
where $\epsilon $\ is the dielectric constant of the medium. The wave  
equation for the electric field resulting from the Maxwell equations is  
  
\begin{equation}  
-\nabla \times (\nabla \times \mathbf{E})-\epsilon \ddot{\mathbf{E}}=\ddot{%
\mathbf{P}},  \label{wave}  
\end{equation}  
By assuming a linear relation between sources and fields, the electric field  
can be written as a spacetime integral 
  
\begin{equation}  
\mathbf{E}(x)=\int d^4x^{\prime }\mathbf{\,}\overleftrightarrow{\mathbf{%
\Gamma }}\mathbf{(}x,x^{\prime }\mathbf{)\,\mathbf{P}(}x^{\prime }\mathbf{),}  
\label{sol}  
\end{equation}  
where $x=(t,\mathbf{r})$, $x'=(t',\mathbf{r}')$ and  
$\overleftrightarrow{\mathbf{\Gamma }}$ is a Green's dyadic, which satisfies 
(\ref{wave}) with a $\delta$-function source. Let 
\begin{equation} 
\overleftrightarrow{\mathbf{\Gamma}}(\mathbf{r},\mathbf{r}',\omega) = 
\int_{-\infty}^{\infty} d\tau \, e^{i \omega \tau} \, 
\overleftrightarrow{\mathbf{\Gamma}}(x,x') \, , 
\end{equation} 
where $\tau = t-t'$. From (\ref  
{wave}) and (\ref{sol}), it follows that $\mathbf{\overleftrightarrow{  
\mathbf{\Gamma }}}(\mathbf{r},\mathbf{r}',\omega)$ satisfies the following  
equation:  
  
\begin{equation}  
-\nabla \times (\nabla \times \mathbf{\overleftrightarrow{\mathbf{\Gamma }})+%
}\omega ^{2}\epsilon \mathbf{\overleftrightarrow{\mathbf{\Gamma }}=-}\omega  
^{2}\,\mathbf{\overleftrightarrow{\mathbf{1}}}\delta \mathbf{(\mathbf{r}-%
\mathbf{r}^{\prime }).}  \label{gwave}  
\end{equation} 
 
So far, the discussion has been purely classical. At this point, 
 Schwinger {\it et al} \cite{one} use source theory to identify the Green's 
dyadic $\overleftrightarrow{\mathbf{\Gamma }}$ with an ``effective product of 
electric fields'' 
  
\begin{equation}  
\frac{i}{\hbar } \langle E_{j} 
(\mathbf{r})\,E_{k}(\mathbf{r}^{\prime })\rangle=\Gamma _{jk}(%
\mathbf{r},\mathbf{r}^{\prime },\omega ).  \label{ef1}  
\end{equation}  
We can interpret this as the Fourier transform of the electric field 
correlation function. From the Maxwell equation  
$\nabla \times \mathbf{E=-\dot{B}}$\textbf{,}  
one finds the corresponding expression for the magnetic field:  
  
\begin{equation}  
\frac{i}{\hbar }\langle B_{j}(\mathbf{r})\,B_{k}(\mathbf{r}^{\prime })\rangle 
=\epsilon  
_{jlm}\epsilon _{knp}(\nabla _{l}\nabla _{n'}/\omega ^{2}\,)\Gamma _{mp}(%
\mathbf{r},\mathbf{r}^{\prime },\omega ).  \label{mf1}  
\end{equation} 
Note that $\hbar $ makes its first appearance in these expressions. 
These expressions can be identified with the vacuum expectation values 
of products of field operators, which appear in the more conventional 
field theory approach to quantization of the electromagnetic field. From 
now onward, we revert to units in which $\hbar=1$.   
 In order to calculate the field correlation functions, one needs to find the  
Green's function $\mathbf{\Gamma }$ occurring in (\ref{gwave}). This 
amounts to solving a classical boundary value problem. 
  
The interfaces between the media are chosen to be perpendicular to the $z$  
direction, so for now it will only matter that the dielectric constant  
changes in the $z$ direction only. Therefore, it is convenient to introduce  
a transverse spatial Fourier transform  
  
\begin{equation}  
\mathbf{\overleftrightarrow{\mathbf{\Gamma }}(\mathbf{r},\mathbf{r}^{\prime  
},}\omega \mathbf{)=\int }d\mathbf{\mathbf{k}_{\bot }\;}\frac{1}{(2\pi )^{2}}%
e^{i\mathbf{k}_{\bot }(\mathbf{r}-\mathbf{r}^{\prime })_{\bot }}\mathbf{%
\overleftrightarrow{\mathbf{\Gamma }}(}z,z^{\prime }\mathbf{,\mathbf{k}%
_{\bot },}\omega \mathbf{),}  \label{gamf}  
\end{equation}  
where the vector $\mathbf{k}_{\perp }$ can be chosen to point along the $+x$  
axis ($k=\left| \mathbf{k}_{\perp }\right| $).  
  
Some components of $\mathbf{\overleftrightarrow{\mathbf{\Gamma }}}$are found  
to be \cite{one}  
  
\begin{mathletters}  
\label{Gcomps}  
\begin{eqnarray}  
\Gamma _{xx} &=&-\frac{1}{\epsilon }\delta (z-z^{\prime })+\frac{1}{\epsilon   
}\frac{\partial }{\partial z}\frac{1}{\epsilon ^{\prime }}\frac{\partial }{%
\partial z^{\prime }}g^{B},  \label{Gxx} \\  
\Gamma _{yy} &=&\omega ^{2}g^{E},  \label{Gyy} \\  
\Gamma _{zz} &=&-\frac{1}{\epsilon }\delta (z-z^{\prime })+\frac{k^{2}}{%
\epsilon \epsilon ^{\prime }}g^{B},  \label{Gzz} \\  
\Gamma _{xz} &=&i\frac{k}{\epsilon \epsilon ^{\prime }}\frac{\partial }{%
\partial z}g^{B},  \label{Gxz} \\  
\Gamma _{zx} &=&-i\frac{k}{\epsilon \epsilon ^{\prime }}\frac{\partial }{%
\partial z^{\prime }}g^{B},  \label{Gzx}  
\end{eqnarray}  
where $\epsilon ^{\prime }=\epsilon (z^{\prime })$, and $g^{E}$, the  
``transverse electric'', and $g^{B}$, the ``transverse magnetic'' Green's  
functions satisfy  
  
\end{mathletters}  
\begin{mathletters}  
\label{gegh}  
\begin{eqnarray}  
\left[ -\frac{\partial ^{2}}{\partial z^{2}}+k^{2}-\omega ^{2}\epsilon %
\right] g^{E}(z,z^{\prime }) &=&\delta (z-z^{\prime }),  \label{ge} \\  
\left[ -\frac{\partial }{\partial z}\frac{1}{\epsilon }\frac{\partial }{%
\partial z}+\frac{k^{2}}{\epsilon }-\omega ^{2}\right] g^{B}(z,z^{\prime })  
&=&\delta (z-z^{\prime }).  \label{gh}  
\end{eqnarray}  
By introducing the quantity  
  
\end{mathletters}  
\begin{equation}  
\kappa ^2=k^2-\omega ^2\epsilon ,  \label{ka}  
\end{equation}  
(\ref{gegh}) can be written as:  
  
\begin{mathletters}  
\label{gegheq}  
\begin{eqnarray}  
\left[ -\frac{\partial ^{2}}{\partial z^{2}}+\kappa ^{2}\right]  
g^{E}(z,z^{^{\prime }}) &=&\delta (z-z^{\prime }),  \label{geq} \\  
\left[ -\frac{\partial }{\partial z}\frac{1}{\epsilon }\frac{\partial }{%
\partial z}+\frac{\kappa ^{2}}{\epsilon }\right] g^{B}(z,z^{\prime })  
&=&\delta (z-z^{\prime }).  \label{gheq}  
\end{eqnarray}  
  
So, in order to find the field correlation functions  
as defined in (\ref{ef1}) and (\ref{mf1}) in  
a given situation, one needs to solve these equations with the appropriate  
boundary conditions. We consider here two cases.  
 
\section{\textbf{One Interface Case}}  
 \label{sec:one} 
  
We now specialize the above discussion to a situation in which the  
inhomogeneity of the dielectric constant is due to a plane interface  
separating a dielectric substance from a vacuum:  
  
\end{mathletters}  
\begin{eqnarray}  
z>0:\,\,\epsilon (z) &=&1, \nonumber \\  
z<0:\,\,\epsilon (z) &\equiv &\epsilon _{d}.  
\end{eqnarray}  
Here $\epsilon_{d}$ is a function of frequency, but not of position. 
  
\subsection{\textit{Boundary Conditions}}  
  
In solving (\ref{geq}) and (\ref{gheq}), we use the following boundary  
conditions. At $z=z^{\prime }$, $g$ is continuous but the derivative is  
discontinuous at this point \cite{two}:  
  
\begin{equation}  
\frac{\partial g}{\partial z}\biggl| _{z\rightarrow z_{-}^{\prime  
}}^{z\rightarrow z_{+}^{\prime }}=-1.  \label{delta}  
\end{equation}  
At the boundary ($z=0$) we use the conditions for continuity of $E_{x}$, $%
E_{y}$, $\epsilon E_{z}$, and $B_{i}$. The first three, as seen from (\ref  
{ef1}), imply the continuity of $\Gamma _{xx}$, $\Gamma _{yy}$, and $%
\epsilon \Gamma _{zz}$ and subsequently, from (\ref{Gcomps}), the continuity  
of $g^{E}$, $g^{B}$, and  
  
\begin{equation*}  
\frac{1}{\epsilon }\frac{\partial }{\partial z}\frac{1}{\epsilon ^{\prime }}%
\frac{\partial }{\partial z^{\prime }}g^{B}.  
\end{equation*}   
The continuity of $B_{x}\,$implies that of $\nabla _{z}\nabla _{z^{\prime  
}}\Gamma _{yy}\,$, as seen from Eq. (\ref{HiHi}), which is given below. From  
this, using (\ref{ef1}) and (\ref{Gyy}), we deduce the continuity of $%
\partial g^{E}/\partial z$.  
  
The solutions $g^{E}$ and $g^{B}$ in the vacuum region have the form  
  
\begin{mathletters}  
\label{ONEgegh}  
\begin{eqnarray}  
g^{E} &=&\frac{e^{-\kappa _{0}\mid z-z^{\prime }\mid }+re^{-\kappa  
_{0}(z+z^{\prime })}}{2\kappa _{0}},  \label{onege} \\  
g^{B} &=&\frac{e^{-\kappa _{0}\mid z-z^{\prime }\mid }+r^{\prime }e^{-\kappa  
_{0}(z+z^{\prime })}}{2\kappa _{0}},  \label{onegh}  
\end{eqnarray}  
where   
\end{mathletters}  
\begin{mathletters}  
\label{Rcoefs}  
\begin{eqnarray}  
r &\equiv &\frac{\kappa _{0}-\kappa _{1}}{\kappa _{0}+\kappa _{1}}  
\label{rcoef} \\  
r^{\prime } &\equiv &\frac{\kappa _{0}\epsilon _{d}-\kappa _{1}}{\kappa  
_{0}\epsilon _{d}+\kappa _{1}}.  \label{rprime}  
\end{eqnarray}  
Here $\kappa _{0}$ and $\kappa _{1}$ represent the quantity $\kappa $ as  
defined in (\ref{ka}) for the vacuum region $(\epsilon =1)$ , and for the  
dielectric half-space region $(\epsilon = \epsilon _{d})$ ,  
respectively, and $r$ and $r^{\prime }$ can be identified as reflection  
coefficients for two polarization states, $\bot $ and $\Vert $ respectively,  
corresponding to electric field vector being perpendicular or parallel to  
the plane of incidence of an linearly polarized electromagnetic wave \cite  
{two}.  
  
\subsection{\textit{The Electric Field}}  
  
Using (\ref{ef1}), we write the formal expectation value of the  
square of the electric field at coincident points as  
  
\end{mathletters}  
\begin{equation}  
\left\langle E^{2}\right\rangle_f =-i\int_{-\infty }^{\infty }d\omega \frac{1}{%
2\pi }\int_{0}^{\infty }dk\,k\;\frac{1}{2\pi }\Gamma _{kk} 
=-\frac{i}{2\pi^2}\int_{0}^{\infty }d\omega  
\int_{0}^{\infty }dk\,k\;\Gamma _{kk}.  \label{esquare}  
\end{equation}  
In the second step, we assumed that the integrand is an even function of 
$\omega$. By complex rotation $(\omega \rightarrow i\zeta )$, this becomes:  
  
\begin{equation}  
\left\langle E^{2}\right\rangle_f =\frac{1}{2\pi ^{2}}\int_{0}^{\infty }d\zeta  
\int_{0}^{\infty }dk\,k\;\Gamma _{kk}.  \label{erot}  
\end{equation}  
Note from (\ref{ka}) that $\kappa^2 > 0$ when $\omega$ is imaginary. 
By means of (\ref{Gcomps}), all of the components of $\overleftrightarrow{%
\mathbf{\Gamma }}$ in a given region can be written in terms of $\Gamma  
_{xx} $ and $\Gamma _{yy}$:  
  
\begin{eqnarray}  
\Gamma _{xz}(z,z^{\prime }) &=&\frac{ik}{\kappa ^{2}}\frac{\partial }{%
\partial z^{\prime }}\,\Gamma _{xx}(z,z^{\prime }),  \notag \\  
\Gamma _{zx}(z,z^{\prime }) &=&-\frac{ik}{\kappa ^{2}}\frac{\partial }{%
\partial z}\,\Gamma _{xx}(z,z^{\prime }),  \label{gamas} \\  
\Gamma _{zz}(z,z^{\prime }) &=&\frac{k^{2}}{(\kappa ^{2})}\frac{\partial }{%
\partial z}\frac{\partial }{\partial z^{\prime }}\,\Gamma _{xx}(z,z^{\prime  
})+\frac{\omega ^{2}}{\kappa ^{2}}\delta (z-z^{\prime }).  \notag  
\end{eqnarray}  
By taking the limit $z\rightarrow z^{\prime }$, and thus omitting the delta  
function, $\Gamma _{kk}$ becomes 
  
\begin{equation}  
\Gamma _{kk}=\Gamma _{xx}+\Gamma _{yy}+\frac{k^{2}}{(\kappa ^{2})^{2}}\nabla  
_{z}\nabla _{z^{\prime }}\Gamma _{xx},  \label{gkk}  
\end{equation}  
or by (\ref{Gcomps}), using $\epsilon =1$,  
  
\begin{eqnarray}  
\Gamma _{kk} &=&\omega ^{2}g^{E}+\nabla _{z}\nabla _{z^{\prime }}g^{B}+\frac{%
k^{2}}{(\kappa ^{2})^{2}}\nabla _{z}\nabla _{z^{\prime }}(\nabla _{z}\nabla  
_{z^{\prime }}g^{B})  \notag \\  
&=&\omega ^{2}g^{E}+(k^{2}+\nabla _{z}\nabla _{z^{\prime }})g^{B}.  
\label{gkk1}  
\end{eqnarray}  
Using (\ref{ka}) and (\ref{ONEgegh}), this becomes  
  
\begin{equation}  
\Gamma _{kk}=\frac{\omega ^{2}}{\kappa }+\frac{1}{2\kappa }\left[ \omega  
^{2}r+\left( 2k^{2}-\omega ^{2}\right) r^{\prime }\right] e^{-2\kappa z}.  
\end{equation} 
 
Equation~(\ref{erot}) gives a formal expectation value only, because the 
integral is divergent. However, the divergence comes only from the  
${\omega ^{2}}/{\kappa }$ term in $\Gamma _{kk}$ and is independent of 
$z$. It is the usual empty space vacuum divergence. We will henceforth 
drop this term and denote the resulting finite expectation value by 
$\left\langle E^{2}\right\rangle $. The renormalization results 
in a quantity which vanishes at large distances from the interface: 
$\left\langle E^{2}\right\rangle \rightarrow 0$ as $z\rightarrow \infty $, 
which amounts to finding the difference in $\left\langle E^{2}\right\rangle $  
with the boundary and without it. Thus we find  
  
\begin{equation}  
\left\langle E^{2}\right\rangle =\frac{1}{4\pi ^{2}}\int_{0}^{\infty }d\zeta  
\int_{0}^{\infty }dk\,\frac{k}{\kappa }\left[ -\zeta ^{2}r+\left(  
2k^{2}+\zeta ^{2}\right) r^{\prime }\right] e^{-2\kappa z}.  \label{oneEF}  
\end{equation}  
  
\subsection{\textit{The Magnetic Field}}  
  
Now we compute the expectation value of the magnetic field. Using Eq. (\ref  
{mf1}), we find  
  
\begin{eqnarray}  
i\langle B_{x}(\mathbf{r})\,B_{x}(\mathbf{r}^{\prime })\rangle  
&=&\frac{1}{\omega ^{2}}%
\left( \nabla _{z}\nabla _{z^{\prime }}\Gamma _{yy}-\nabla _{y}\nabla  
_{z^{\prime }}\Gamma _{zy}-\nabla _{z}\nabla _{y^{\prime }}\Gamma  
_{yz}+\nabla _{y}\nabla _{y^{\prime }}\Gamma _{zz}\right) ,  \notag \\  
i\langle B_{y}(\mathbf{r})B_{y}(\mathbf{r}^{\prime })\rangle 
 &=&\frac{1}{\omega ^{2}}\left(  
\nabla _{z}\nabla _{z^{\prime }}\Gamma _{xx}-\nabla _{x}\nabla _{z^{\prime  
}}\Gamma _{zx}-\nabla _{z}\nabla _{x^{\prime }}\Gamma _{xz}+\nabla  
_{x}\nabla _{x^{\prime }}\Gamma _{zz}\right) ,  \label{HiHi} \\  
i\langle B_{z}(\mathbf{r})\,B_{z}(\mathbf{r}^{\prime })\rangle 
 &=&\frac{1}{\omega ^{2}}%
\left( \nabla _{y}\nabla _{y^{\prime }}\Gamma _{xx}-\nabla _{y}\nabla  
_{x^{\prime }}\Gamma _{xy}-\nabla _{x}\nabla _{y^{\prime }}\Gamma  
_{yx}+\nabla _{x}\nabla _{x^{\prime }}\Gamma _{yy}\right) .  \notag  
\end{eqnarray}  
  
   From the definition of $\mathbf{k}_{\perp }$, it follows that all  
derivatives in $y$ vanish, so we can write the sum of the above terms as  
  
\begin{align}  
i\langle B_{i}(\mathbf{r})B_{i}(\mathbf{r}^{\prime })\rangle 
& =\frac{1}{\omega ^{2}}\left(  
\nabla _{z}\nabla _{z^{\prime }}\Gamma _{yy}+\nabla _{z}\nabla _{z^{\prime  
}}\Gamma _{xx}-\right.  \notag \\  
& \left. \nabla _{x}\nabla _{z^{\prime }}\Gamma _{zx}-\nabla _{z}\nabla  
_{x^{\prime }}\Gamma _{xz}+\nabla _{x}\nabla _{x^{\prime }}\Gamma  
_{zz}+\nabla _{x}\nabla _{x^{\prime }}\Gamma _{yy}\right) .  \label{hii}  
\end{align}  
Using (\ref{gamas}), we have  
  
\begin{eqnarray}  
\nabla _{x}\nabla _{x^{\prime }}\Gamma _{zz}(\mathbf{r},\mathbf{r}^{\prime  
},\omega ) &=&\int \frac{d\mathbf{k}_{\perp }}{(2\pi )^{2}}\;\frac{%
(k^{2})^{2}}{(\kappa ^{2})^{2}}\;\nabla _{z}\nabla _{z^{\prime }}\Gamma  
_{xx},  \notag \\  
\nabla _{z}\nabla _{x^{\prime }}\Gamma _{xz}(\mathbf{r},\mathbf{r}^{\prime  
},\omega ) &=&\int \frac{d\mathbf{k}_{\perp }}{(2\pi )^{2}}\;\frac{k^{2}}{%
\kappa ^{2}}\;\nabla _{z}\nabla _{z^{\prime }}\Gamma _{xx,}  \label{deltas}  
\\  
\nabla _{x}\nabla _{z^{\prime }}\Gamma _{zx}(\mathbf{r},\mathbf{r}^{\prime  
},\omega ) &=&\int \frac{d\mathbf{k}_{\perp }}{(2\pi )^{2}}\;\frac{k^{2}}{%
\kappa ^{2}}\;\nabla _{z}\nabla _{z^{\prime }}\Gamma _{xx}.  \notag  
\end{eqnarray}  
This leads to  
  
\begin{equation}  
i\langle B(x)\,B(x^{\prime })\rangle 
=\int \frac{d\omega }{2\pi }\int \frac{d\mathbf{k}%
_{\perp }}{(2\pi )^{2}}\left[ \frac{1}{\omega ^{2}}\left( k^{2}+\nabla  
_{z}\nabla _{z^{\prime }}\right) \Gamma _{yy}(z,z^{\prime })+\frac{\omega  
^{2}}{\kappa ^{4}}\nabla _{z}\nabla _{z^{\prime }}\Gamma _{xx}(z,z^{\prime })%
\right] .  \label{hfur}  
\end{equation}  
Using (\ref{Gcomps}), this becomes  
  
\begin{equation}  
i\langle B(x)\,B(x^{\prime })\rangle 
=\int \frac{d\omega }{2\pi }\int \frac{d\mathbf{k}%
_{\perp }}{(2\pi )^{2}}\left[ \left( k^{2}+\nabla _{z}\nabla _{z^{\prime  
}}\right) g^{E}(z,z^{\prime })+\omega ^{2}g^{B}(z,z^{\prime })\right] .  
\label{hgegh}  
\end{equation}  
Following the same procedure as used above in calculating $\left\langle  
E^{2}\right\rangle $, we find the finite mean squared magnetic field to be  
  
\begin{equation}  
\left\langle B^{2}\right\rangle =\frac{1}{4\pi ^{2}}\int_{0}^{\infty }d\zeta  
\int_{0}^{\infty }dk\;\frac{k}{\kappa }\left[ \left( 2k^{2}+\zeta  
^{2}\right) r-\zeta ^{2}r^{\prime }\right] e^{-2\kappa z}.  \label{oneMF}  
\end{equation}  
Note from (\ref{oneEF}) and (\ref{oneMF}) that 
$\langle E^{2}\rangle \leftrightarrow \langle B^{2}\rangle$     
under interchange of $r$ and $r'$. Now, the mean energy density can  
be calculated as  
  
\begin{equation}  
U=\frac{1}{2}\left( \left\langle E^{2}\right\rangle +\left\langle  
B^{2}\right\rangle \right) .  \label{etensor}  
\end{equation}  
Using (\ref{oneEF}) and (\ref{oneMF}), this becomes  
  
\begin{equation}  
U=\frac{1}{4\pi ^{2}}\int_{0}^{\infty }d\zeta \int_{0}^{\infty }dk\,\frac{%
k^{3}}{\kappa }\left( r+r^{\prime }\right) e^{-2\kappa z}.  \label{energy1}  
\end{equation}  
We can write $U$ in a form more convenient for numerical calculation by  
introducing polar coordinates $u$ and $\theta $ ($\zeta =u\cos \theta ,$ $%
k=u\sin \theta $):  
  
\begin{equation}  
U=\frac{1}{4\pi ^{2}}\int_{0}^{\infty }du\,u^{3}\int_{0}^{\frac{\pi }{2}%
}d\theta \,(\sin \theta )^{3}\left( r+r^{\prime }\right) e^{-2uz}.  
\label{energy11polar}  
\end{equation}  
We use the Drude model for the dielectric function  
  
\begin{equation}  
\epsilon _{d}(\omega )=1-\frac{\omega _{p}^{2}}{\omega ^{2}},  \label{drude}  
\end{equation}  
where $\omega _{p}$ is the plasma frequency. From (\ref{rcoef}), (\ref  
{rprime}), and (\ref{drude}), we find  
  
\begin{mathletters}  
\label{Rpolars}  
\begin{eqnarray}  
r &=&\frac{u-\sqrt{u^{2}+\omega _{p}^{2}}}{u+\sqrt{u^{2}+\omega _{p}^{2}}},  
\label{Rpolar} \\  
r^{\prime } &=&\frac{u^{2}(\cos \theta )^{2}+\omega _{p}^{2}-u(\cos \theta  
)^{2}\sqrt{u^{2}+\omega _{p}^{2}}}{u^{2}(\cos \theta )^{2}+\omega  
_{p}^{2}+u(\cos \theta )^{2}\sqrt{u^{2}+\omega _{p}^{2}}}.  \label{R'polar}  
\end{eqnarray}  
By a substitution ($\cos (\theta )\rightarrow t$), U becomes  
  
\end{mathletters}  
\begin{equation}  
U=\frac{1}{4\pi ^{2}}\int_{0}^{\infty }du\,u^{3}\left[  
\int_{0}^{1}dt\,(1-t^{2})\left( r+r^{\prime }\right) \right] e^{-2uz}.  
\label{energy1num}  
\end{equation}  
By the same coordinate transform, (\ref{oneEF}) and (\ref{oneMF}) become  
  
\begin{mathletters}  
\label{EFMF1}  
\begin{eqnarray}  
\left\langle E^{2}\right\rangle &=&\frac{1}{4\pi ^{2}}\int_{0}^{\infty  
}du\,u^{3}\left\{ \int_{0}^{1}dt\;\left[ -t^{2}r+(2-t^{2})r^{\prime }\right]  
\right\} e^{-2uz},  \label{EF1} \\  
\left\langle B^{2}\right\rangle &=&\frac{1}{4\pi ^{2}}\int_{0}^{\infty  
}du\,u^{3}\left\{ \int_{0}^{1}dt\;\left[ (2-t^{2})r-t^{2}r^{\prime }\right]  
\right\} e^{-2uz}.  \label{MF1}  
\end{eqnarray}

\begin{figure}  
\begin{center}  
\leavevmode\epsfysize=6cm\epsffile{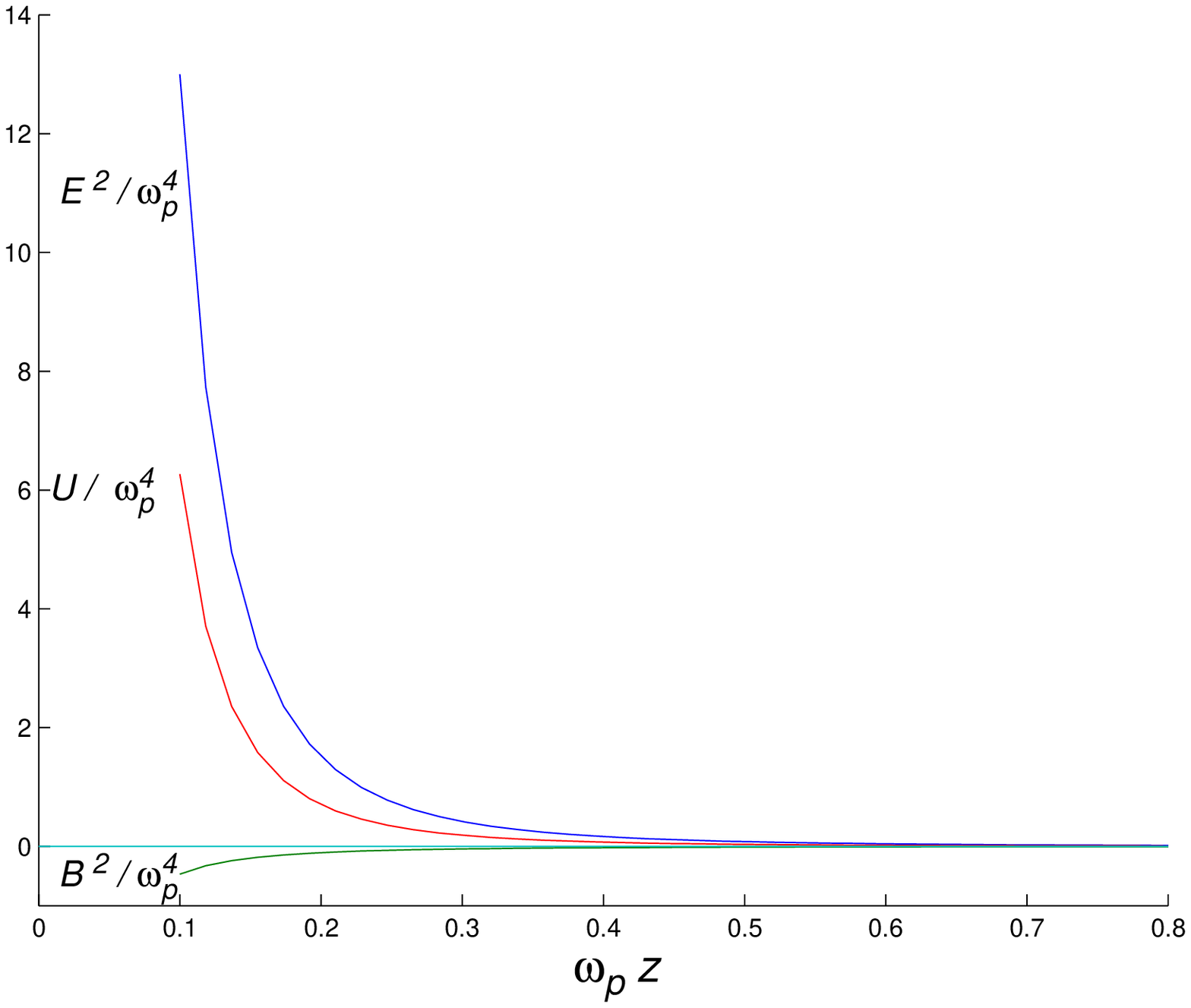}  
\label{Figure 1}  
\end{center}  
\begin{caption}[] 
   
The expectations of the squares of electric field, magnetic field,   
and energy density near the dielectric half-space are illustrated.  
\end{caption}  
\end{figure}

The plot for $\left\langle E^{2}\right\rangle $ and $\left\langle  
B^{2}\right\rangle ,$ as well as $U$ is shown in Figure 1. As we can see  
from the figure, the energy density is positive. Now we consider some  
limiting cases.  
  
\subsection{\textit{The Fields near the Interface}}  
  
To see how $U$ behaves for small $z$ (large $u$), we first perform the $t$  
integration in (\ref{energy1num}), which can be done analytically, and then  
Taylor expand the resulting expression in the brackets in powers of $u^{-1}$. 
That is, we are expanding all of the integrand except for the exponential 
factor.  
To the leading order we find:  
  
\end{mathletters}  
\begin{equation}  
U\sim \frac{\sqrt{2}\omega _{p}}{64\pi }\frac{1}{z^{3}}.  \label{enclose}  
\end{equation}  
The asymptotic behavior of the mean squared fields (\ref{EF1}), (\ref{MF1})  
in this limit is  
  
\begin{mathletters}  
\label{EHclose}  
\begin{eqnarray}  
\left\langle E^{2}\right\rangle &\sim &\frac{\sqrt{2}\omega _{p}}{32\pi }%
\frac{1}{z^{3}},  \label{eclose} \\  
\left\langle B^{2}\right\rangle &\sim &-\frac{5\omega _{p}^{2}}{96\pi }%
\frac{1}{z^{2}}.  \label{hclose}  
\end{eqnarray}  
We see that $\left\langle E^{2}\right\rangle $ dominates over $\left\langle  
B^{2}\right\rangle $, so that $U\approx 1/2\left\langle E^{2}\right\rangle ;$  
this is due fact that the leading order in the expression in braces in (\ref  
{EF1}) is $\propto u^{-1}$ as compared to $u^{-2}$ in (\ref{MF1}). If we  
compare these expressions to ones that would result if dispersion were not  
included in the calculation, it can be seen from (\ref{EFMF1}) that in this  
case $\left\langle E^{2}\right\rangle \propto z^{-4}$, and same for $%
\left\langle B^{2}\right\rangle $ (see also \cite{Lang}). As seen from (\ref  
{EHclose}), the inclusion of dispersion in the calculation reduces the power  
in $z$ up to two orders, but it does not remove the singularity of the  
results at $z=0$, as might be naively expected.  
 
After more careful consideration, it is not surprising that dispersion 
alone is insufficient to render the results finite at the interface. The  
integrals for  $\left\langle E^{2}\right\rangle $ and $\left\langle  
B^{2}\right\rangle $ at $z=0$ diverge quartically in a frequency cutoff. 
However, in general dielectric functions go as  
\begin{equation}  
\epsilon(\omega) \sim 1 + O(\omega^{-2}) 
\end{equation} 
as $\omega \rightarrow \infty$. Thus the reflection coefficients will 
go to zero no faster than $\omega^{-2}$, leaving the integrals  
quadratically divergent. This argument explains why  
$\left\langle B^{2}\right\rangle \propto z^{-2}$ for small $z$, 
but understanding 
the behavior of $\left\langle E^{2}\right\rangle $ requires examining the 
dependence of the reflection coefficients $r$ and $r'$ upon the transverse 
momentum $k$. In fact the contribution of the coefficient $r$, which describes  
modes with the polarization vector perpendicular to the plane of incidence, 
does go as $z^{-2}$. This coefficient depends only upon frequency,  and falls 
as $\omega^{-2}$ for large $\omega$, as can be seen 
 from (\ref{Rpolar}). The coefficient $r'$ describes modes with the polarization 
vector parallel to the plane of incidence, and goes to one as $\theta 
\rightarrow \pi/2$ (corresponding to grazing incidence) for all frequencies. 
It is this behavior which leads to the $z^{-3}$ singularity in 
$\left\langle E^{2}\right\rangle $ and hence in $U$. (The role of cutoffs 
for the quantized electromagnetic field in dielectrics has been discussed 
in more detail by Candelas \cite{Candelas}. Barton \cite{Barton02} has recently 
emphasized the fact that dispersion alone will not remove all divergences.) 
 
The divergence of $U$ is not  
considered to be physical, but as resulting from the idealization of the wall as  
a perfectly smooth surface. One way of removing this singularity is to 
allow the position of the boundary to fluctuate \cite{Ford&Svaiter}. 
It seems plausible that such effects as surface roughness, or the atomic 
nature of matter on small scales can also introduce a physical cutoff 
that makes the mean squared fields and the energy density finite everywhere.  
  
\subsection{\textit{Case of a Perfect Conductor}}  
  
Now we consider the limit $\epsilon \rightarrow \infty $. In this limit, as  
seen from (\ref{ka}), (\ref{rcoef}) and (\ref{rprime}), $r\rightarrow -1$,  
and $r^{\prime }\rightarrow 1$ . Equation~(\ref{energy1}) implies that  
$U$ becomes zero, as expected, and (\ref{EF1}) and (\ref{MF1}) give:  
  
\end{mathletters}  
\begin{eqnarray}  
\left\langle E^{2}\right\rangle &\sim &\frac{3}{16\pi ^{2}}\frac{1}{z^{4}}%
,  \label{econd} \\  
\left\langle B^{2}\right\rangle &\sim &-\frac{3}{16\pi ^{2}}\frac{1}{z^{4}%
}.  \label{hcond}  
\end{eqnarray}  
 These well-known results are consistent with the asymptotic Casimir-Polder  
potential \cite{polder}:  
  
\begin{equation}  
V_{CP}\sim -\frac{3}{32\pi ^{2}}\frac{\alpha _{0}}{z^{4}}=-\frac{1}{2}\alpha  
_{0}\left\langle E^{2}\right\rangle ,  \label{VPolder}  
\end{equation}  
where $\alpha _{0}$ is the static polarizability of an atom near the  
interface.  
  
\section{\textbf{The energy density between dielectrics}}  
\label{sec:two} 
  
In this section we calculate the energy density in a vacuum region of width   
$a$ between two dielectric half-spaces. We define the dielectric constant as:  
  
\begin{eqnarray}  
0<z<a:\,\,\epsilon &=&1, \nonumber \\  
z<0\,\ \mathrm{and\,}\,z>a:\,\,\epsilon &= &\epsilon _{d}.  
\end{eqnarray}  
In the vacuum region, $g^{E}$ occurring in (\ref{geq}) has the form  
  
\begin{align}  
g^{E}& =-\frac{1}{2 \kappa _{0}}\left\{ \frac{\exp (-\kappa _{0}\mid  
z-z^{\prime }\mid )+r^{2}\exp (-2\kappa _{0}a)\exp (\kappa _{0}\mid  
z-z^{\prime }\mid )}{r^{2}\exp (-2\kappa _{0}a)-1}+\right.  \notag \\  
& \left. +\frac{r\left[ \exp (-\kappa _{0}(z+z^{\prime }))+\exp (-2\kappa  
_{0}a)\exp (\kappa (z+z^{\prime }))\right] }{r^{2}\exp (-2\kappa _{0}a)-1}%
\right\} ,  \label{gesol}  
\end{align}  
and $g^{B}$ has the same form as $g^{E,}$ with $r$ and $r'$ interchanged,  
where $r$ and $r^{\prime }$ are defined in (\ref{rcoef}) and (\ref{rprime}).  
We can now calculate the electric and magnetic fields in this region.  
  
\subsection{\textit{The Electric Field}}  
  
The first and the second term on the right hand side of (\ref{gkk1}) 
 for the present case, in the limit $  
z\rightarrow z^{\prime }$, are  
  
\begin{eqnarray}  
\omega ^{2}g^{E} &=&\frac{\omega ^{2}}{2\kappa }\frac{1+r^{2}e^{-2\kappa a}}{%
1-r^{2}e^{-2\kappa a}}+\frac{\omega ^{2}}{2\kappa }\frac{r(e^{-2\kappa  
z}+e^{-2\kappa a}e^{2\kappa z})}{1-r^{2}e^{-2\kappa a}}  \label{gkkone} \\  
(k^{2}+\nabla _{z}\nabla _{z^{\prime }})g^{B} &=&\frac{\omega ^{2}}{2\kappa }%
\frac{1+r^{\prime }{}^{2}e^{-2\kappa a}}{1-r^{\prime }{}^{2}e^{-2\kappa a}}+%
\frac{k^{2}+\kappa ^{2}}{2\kappa }\frac{r^{\prime }(e^{-2\kappa  
z}+e^{-2\kappa a}e^{2\kappa z})}{1-r^{\prime }{}^{2}e^{-2\kappa a}},  
\label{gkktwo}  
\end{eqnarray}  
so that $\Gamma _{kk}$ becomes  
  
\begin{align}  
\Gamma _{kk}& =\frac{\omega ^{2}}{\kappa }+\frac{\omega ^{2}}{\kappa } 
\left(\frac{  
r^{2}}{e^{2\kappa a}-r^{2}}+\frac{r^{\prime }{}^{2}}{e^{2\kappa a}-r^{\prime  
}{}^{2}}\right)+  \notag \\  
& +\frac{1}{\kappa }\left[ \omega ^{2}\frac{r}{1-r^{2}e^{-2\kappa a}}%
+(2k^{2}-\omega ^{2})\frac{r^{\prime }}{1-r^{\prime }{}^{2}e^{-2\kappa a}}%
\right] e^{-\kappa a}\cosh \left[ \kappa (2z-a)\right] .  \label{maki}  
\end{align} 
We again drop the first term on the right hand side of the above expression.   
After introducing polar coordinates and using (\ref{ka}) and (\ref{maki}), (%
\ref{erot}) we find 
  
\begin{align}  
\left\langle E^{2}\right\rangle & =\frac{1}{2\pi ^{2}}\int_{0}^{\infty  
}du\,u^{3}\int_{0}^{\frac{\pi }{2}}d\theta \,\sin (\theta )\left\{ \cos  
^{2}(\theta )\left( \frac{r^{2}}{r^{2}-e^{2ua}}+\frac{r^{\prime }{}^{2}}{%
r^{\prime }{}^{2}-e^{2ua}}\right) +\right.  \notag \\  
& \left. +\left[ -\cos ^{2}(\theta )\frac{r}{1-r^{2}e^{-2ua}}+\left( 1+\sin  
^{2}(\theta )\right) \frac{r^{\prime }}{1-r^{\prime }{}^{2}e^{-2ua}}\right]  
e^{-ua}\cosh \left[ u\left( 2z-a\right) \right] \right\}\, ,  \label{epol}  
\end{align}  
or with ($\cos (\theta )\rightarrow t$)  
  
\begin{align}  
\left\langle E^{2}\right\rangle & =\frac{1}{2\pi ^{2}}\int_{0}^{\infty  
}du\,u^{3}\int_{0}^{1}dt\left\{ t^{2}\left( \frac{r^{2}}{r^{2}-e^{2ua}}+%
\frac{r^{\prime }{}^{2}}{r^{\prime }{}^{2}-e^{2ua}}\right) +\right.  \notag  
\\  
& \left. +\left[ -t^{2}\frac{r}{1-r^{2}e^{-2ua}}+\left( 2-t^{2}\right) \frac{%
r^{\prime }}{1-r^{\prime }{}^{2}e^{-2ua}}\right] e^{-ua}\cosh \left[ u\left(  
2z-a\right) \right] \right\} .  \label{et}  
\end{align}  
  
\subsection{\textit{The Magnetic Field}}  
  
In the same way, (\ref{hgegh}) leads to  
  
\begin{align}  
\left\langle B^{2}\right\rangle & =\frac{1}{2\pi ^{2}}\int_{0}^{\infty  
}du\,u^{3}\int_{0}^{1}dt\left\{ t^{2}\left( \frac{r^{2}}{r^{2}-e^{2ua}}+%
\frac{r^{\prime }{}^{2}}{r^{\prime }{}^{2}-e^{2ua}}\right) +\right.  \notag  
\\  
& \left. +\left[ \left( 2-t^{2}\right) \frac{r}{1-r^{2}e^{-2ua}}-t^{2}\frac{%
r^{\prime }}{1-r^{\prime }{}^{2}e^{-2ua}}\right] e^{-ua}\cosh \left[ u\left(  
2z-a\right) \right] \right\} .  \label{hsquare}  
\end{align}  
This expression differs from the one for $\left\langle E^{2}\right\rangle $  
only in the $z$- dependent term with $r\leftrightarrow r^{\prime }$. A plot  
of $\left\langle E^{2}\right\rangle $ , $\left\langle B^{2}\right\rangle $  
and $U$ is shown in Figure 2 for $\omega _{p}a=200.$ It can be seen in this  
figure that significant, but not complete cancellation occurs between $%
\left\langle E^{2}\right\rangle $ and $\left\langle B^{2}\right\rangle .$

\begin{figure}  
\begin{center}  
\leavevmode\epsfysize=6cm\epsffile{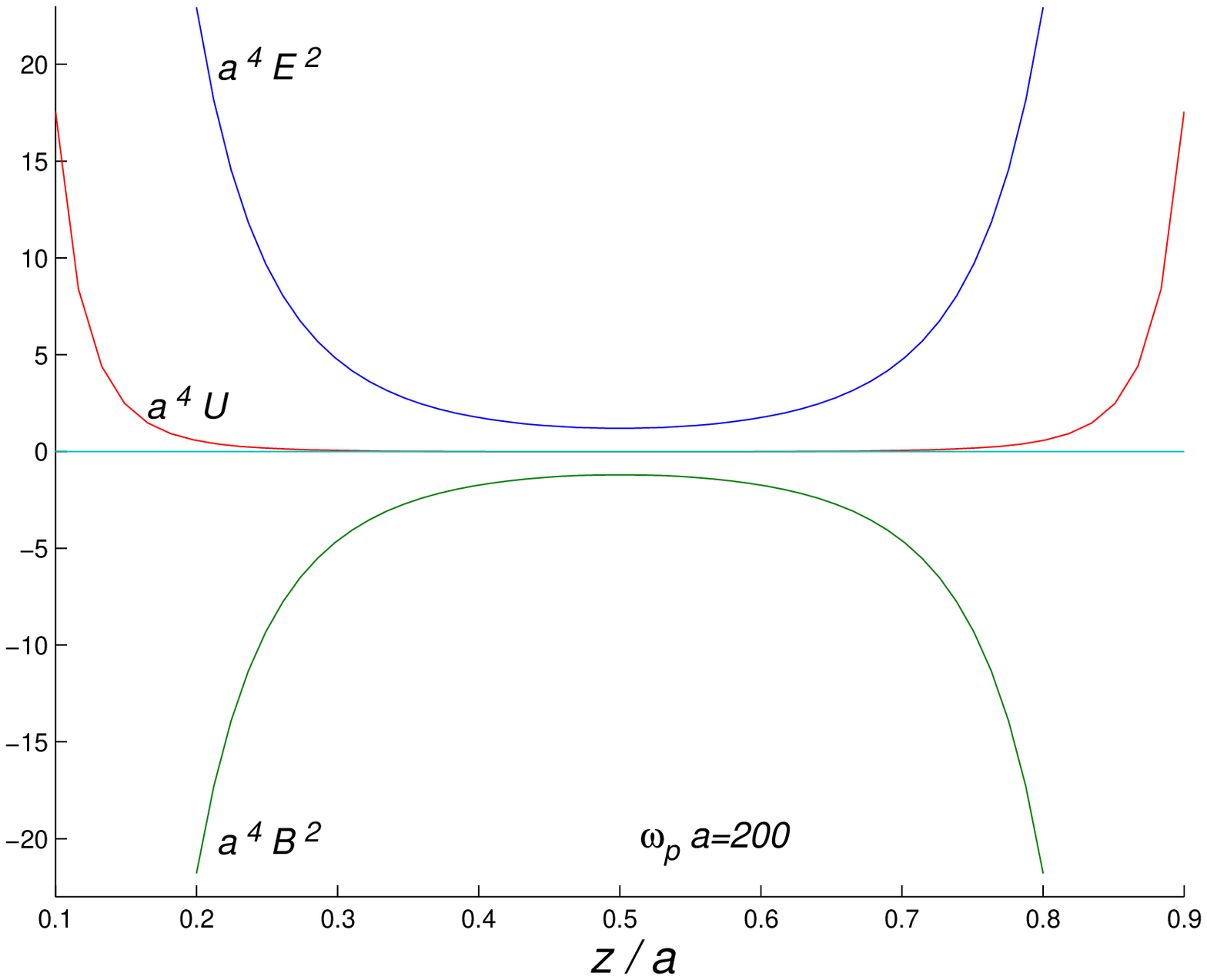}  
\label{Figure 2}  
\end{center}  
\begin{caption}[] 
   
The expectation values of the squared electric and magnetic fields, as well as   
the energy density in the vacuum region, are illustrated   
for $\protect\omega _{p}a=200.$  
\end{caption}  
\end{figure}

\subsection{\textit{The Energy Density}}  
  
Now using (\ref{etensor}), (\ref{et}), and (\ref{hsquare}), the energy  
density in the vacuum region can be calculated as  
  
\begin{align}  
U& =\frac{1}{2\pi ^{2}}\int_{0}^{\infty }du\,u^{3}\int_{0}^{1}dt\left\{  
t^{2}\left( \frac{r^{2}}{r^{2}-e^{2ua}}+\frac{r^{\prime }{}^{2}}{r^{\prime  
}{}^{2}-e^{2ua}}\right) +\right.  \notag \\  
& \left. +\left( 1-t^{2}\right) \left[ \frac{r}{1-r^{2}e^{-2ua}}+\frac{%
r^{\prime }}{1-r^{\prime }{}^{2}e^{-2ua}}\right] e^{-ua}\cosh \left[ u\left(  
2z-a\right) \right] \right\}  \label{energy2}  
\end{align}  
By analyzing this expression we can make some conclusions about the sign of $%
U$. First we note that it is position dependent, and we also note that the  
first term is always negative and the second term is always positive. The  
overall sign of $U$ depends on the choice of $a$ and $\omega _{p}$. As $%
\omega _{p}a$ grows, $U$ at the midpoint decreases, becoming negative for $%
\omega _{p}a\approx 100$, as seen in Figure 3 and Figure 4.  
  

\begin{figure}  
\begin{center}  
\leavevmode\epsfysize=6cm\epsffile{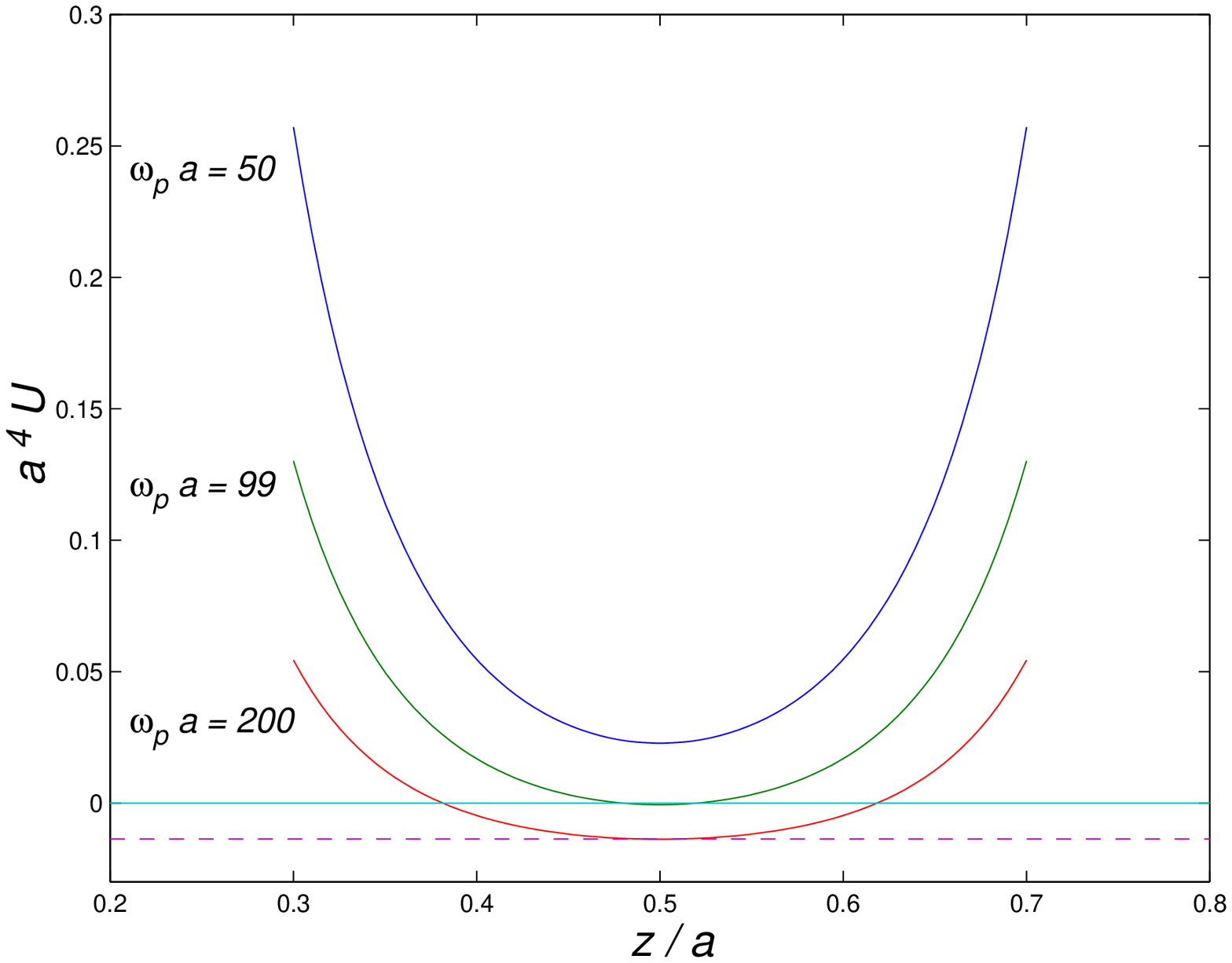}  
\label{Figure 3}  
\end{center}  
\begin{caption}[]  
  
The energy density in the vacuum region between two dielectric   
half-spaces is illustrated for three values of the   
parameter $\protect\omega _{p}a.$ The dashed horizontal line is the energy 
density for the perfectly conducting limit, (\ref{Ucas}). 
\end{caption}  
\end{figure}  
  
In Figure 4 we see how the energy density at the center of the vacuum  
region changes as the product $\omega _{p}a$ increases. It can be seen both  
in Figure 3 and Figure 4 that U approaches the value given in (\ref{Ucas})  
as $\omega _{p}a$ becomes large. The separation at which $U$ becomes 
negative at the center is 
\begin{equation} 
a > a_c = \frac{99}{\omega_{p}} =  
1.3 \mu{\rm m} \left(\frac{14.8 e{\rm V}}{\omega_{p}}\right) \,, 
\end{equation} 
where $14.8 e{\rm V}$ is the plasma frequency of aluminum.  
 
\begin{figure}  
\begin{center}  
\leavevmode\epsfysize=6cm\epsffile{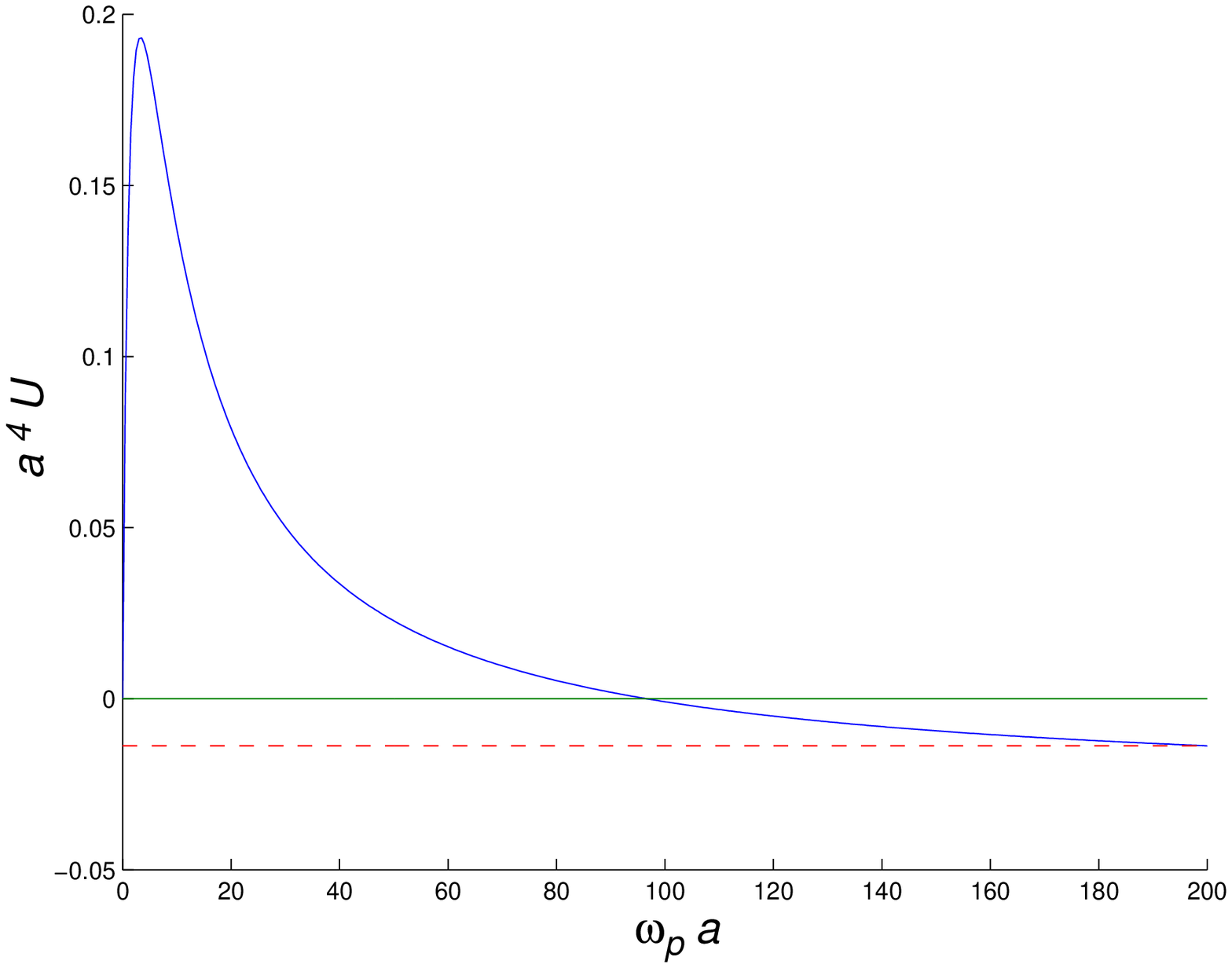}  
\label{Figure 4}  
\end{center}  
\begin{caption}[]  
  
The graph represents the energy density at the center of the gap between   
the two dielectric half-spaces as a function of $\omega_{p} a$.  As   
seen from the graph, the energy density at center becomes negative   
when $\omega_{p} a\approx 99$ or larger. Again the dashed line is 
the perfectly conducting limit. 
\end{caption}  
\end{figure}

\subsection{A Perfect Conductor Case}  
  
In the limit $\omega _{p}\rightarrow \infty $, $r\rightarrow -1$ and $%
r^{\prime }\rightarrow 1.$ Then only the first ($z$-independent) term  
survives in (\ref{energy2}), and we get the familiar result \cite{Brown}:  
  
\begin{equation}  
U=-\frac{\pi ^{2}}{720a^{4}}.  \label{Ucas}  
\end{equation}  
 
It is also interesting to examine the expressions for $\left\langle 
E^{2}\right\rangle $ and $\left\langle B^{2}\right\rangle $ in this limit. 
 After performing the $t$ integration, (\ref{et}) can be written as 
 
\begin{equation} 
\left\langle E^{2}\right\rangle =\frac{1}{2\pi ^{2}}\int_{0}^{\infty 
}du\,u^{3}\left[ \frac{2}{3}\frac{1}{1-e^{2ua}}+\left( \frac{e^{-2u(a-z)}}{%
1-e^{-2ua}}+\frac{e^{-2uz}}{1-e^{-2ua}}\right) \right]\,, 
\end{equation} 
 and after performing the $u$ integration as 
 
\begin{equation} 
\left\langle E^{2}\right\rangle =-\frac{\pi ^{2}}{720a^{4}}+\frac{1}{32\pi 
^{2}a^{4}}\left[ \psi^{(3)}\left(\frac{z}{a}\right)+
\psi^{(3)}\left(1-\frac{z}{a}\right)\right]\,. 
\end{equation} 
 Here $\psi^{(3)}=\frac{d^4}{dz^4} \ln \Gamma(z)$ is the polygamma function 
of order three. It satisfies the 
reflection formula \cite{AS}
\begin{equation} 
\psi^{(3)}\left(\frac{z}{a}\right)+\psi^{(3)}\left(1-\frac{z}{a}\right)
=-\pi \frac{d^{3}}{d(\frac{z}{a})^{3}}\cot \left(\pi \frac{z}{a}\right). 
\end{equation}  
This yields  
\begin{equation} 
\left\langle E^{2}\right\rangle =-\frac{\pi ^{2}}{720a^{4}}+\frac{\pi ^{2}}{%
16a^{4}}\frac{1+2\cos ^{2}(\pi \frac{z}{a})}{\sin ^{4}(\pi \frac{z}{a})}. 
\end{equation}  
In the same way, one finds  
\begin{equation} 
\left\langle B^{2}\right\rangle =-\frac{\pi ^{2}}{720a^{4}}-\frac{\pi ^{2}}{%
16a^{4}}\frac{1+2\cos ^{2}(\pi \frac{z}{a})}{\sin ^{4}(\pi \frac{z}{a})}. 
\end{equation}  
These results are in agreement with those given by previous authors
\cite{Lutken,Barton90}. 
  
\subsection{\textit{Energy density near the Boundary}}  
  
To see how $U$ grows near the interface we note that for small $z$ (large $u$%
), (\ref{energy2}) becomes  
  
\begin{equation}  
U\approx \frac{1}{4\pi ^{2}}\int_{0}^{\infty  
}du\,u^{3}\int_{0}^{1}dt\;\left( 1-t^{2}\right) \left( r+r^{\prime }\right)  
e^{-2uz},  \label{enone}  
\end{equation}  
so it reduces to (\ref{energy1num}), the solution for one interface case. In  
the $z\rightarrow 0$ limit, this expression reduces to (\ref{enclose}).   
By the same reasoning, the expressions for $\left\langle E^{2}\right\rangle $  
and $\left\langle B^{2}\right\rangle $, (\ref{et}) and (\ref{hsquare})  
reduce to (\ref{EF1}) and (\ref{MF1}), respectively, in the small $z$ limit.  
  
\section{Conclusion} 
\label{sec:final} 
 
In this paper we have examined the mean squared fields and the energy density 
in the region between a pair of half-spaces filled with dispersive media. 
We found that these quantities diverge at the boundaries of the media, 
despite the inclusion of dispersion in the calculation. This divergence 
indicates a breakdown of a continuum description in which the dielectric 
function changes discontinuously at the boundary. It also shows that there 
is a positive self-energy density in the region outside of a single plate. 
Nonetheless, we have also found that it is possible for the net energy 
density in the region between the plates to be negative, depending upon the  
plate separation and the plasma frequency of the material involved. 
The existence of an attractive Casimir force is not an indicator of whether the 
energy density at the center of the plates is actually negative or not. 
 
We have found that the energy density at the center becomes negative 
when $\omega_p a > 100$. Thus for fixed plasma frequency $\omega_p$, the 
energy density always becomes negative for sufficiently large separation 
$a$. Of course, in this limit the magnitude of the energy density is also 
becoming small. Similarly, for fixed $a$, the energy density becomes 
negative for sufficiently large $\omega_p$. In the limit that  
$\omega_p \rightarrow \infty$ our results approach the constant negative energy 
density of the perfectly conducting plates. It should not come as a 
surprise that there is a regime of negative energy density. The calculation 
assuming perfect conductivity does have a region of validity so long as 
$\omega_p$ is large and one is not too close to one plate. 
Qualitatively similar behavior has recently been found for the vacuum
energy density near a domain wall \cite{OG}.
 
In this paper, we assumed a particular form for the dielectric function, 
(\ref{drude}), given by the collisionless Drude model. This is a good model 
for many metals especially alkali metals, and is the generic form 
for the dielectric function of all materials at high frequencies. Thus 
taking a different form for $\epsilon_d$ would change the details of 
our results, especially far from an interface, but should lead to the same 
limiting forms near an interface. We have also assumed zero temperature 
throughout this paper. For systems at room temperature, this should be 
a good approximation when the separations are of the order of a few 
micrometers. More generally, one can ignore thermal effects at distances 
small compared to $1/(kT)$.  
 
Contrary to the view expressed by Lamoreaux \cite{lam}, the appearance of  
negative energy density in a quantum field theory is very natural.  
One can easily find quantum states of the free quantized electromagnetic field 
in empty space which have local negative energy densities. A squeezed 
vacuum state is an example \cite{Caves,FGO92}. The energy density of a quantized 
field has to be defined as a difference between that in empty Minkowski  
spacetime, and that in a given state and is no longer positive definite, 
as it was for a classical field. Apart from coupling to gravity, which 
produces extremely small effects, no clear way has been found to directly 
observe the local energy density. In certain limits, the negative energy 
density in a squeezed vacuum state has been shown theoretically \cite{FGO92} 
to produce an effect on the magnetic moment of a spin system. Whether this 
effect could ever be observed, and whether negative Casimir energy density can 
produce similar effects is unknown.

\vspace{0.5cm} 
 
{\bf Acknowledgement:} We would like to thank J.-T. Hsiang 
for valuable discussions. This work was supported in part by the National 
Science Foundation under Grant PHY-9800965.

\ \ \ \ \ \

\end{document}